\documentstyle[12pt]{article}
\newcommand{\Cslash}{\not \!\! C}
\newcommand{\Dslash}{\not \!\! D}
\begin{document}

\makeatletter
\@addtoreset{equation}{section}
\def\theequation{\thesection.\arabic{equation}}
\makeatother

\begin{flushright}{UT-865, 1999}
\end{flushright}
\vskip 0.5 truecm
%\vskip 0.5 truecm

\begin{center}
{\large{\bf  Implications of $Tr\gamma_{5}=0$ in Lattice Gauge 
Theory\footnote{Talk given at Chiral
'99, September 13-18,1999, Taipei, Taiwan(To be published
in the Proceedings)} }}
\end{center}
\vskip .5 truecm
\centerline{\bf Kazuo Fujikawa}
\vskip .4 truecm
\centerline {\it Department of Physics,University of Tokyo}
\centerline {\it Bunkyo-ku,Tokyo 113,Japan}
\vskip 0.5 truecm

\begin{abstract}
We analyze the implications of the relation $Tr\gamma_{5}=0$,
 which is customarily assumed in practical lattice
calculations. On the basis of the finite dimensional 
representations of the Ginsparg-Wilson algebra, it  is shown 
that this relation reflects  the 
species doubling in  lattice theory; topological excitations
associated with species doublers, which have eigenvalue 
$2/a$, contribute 
to $Tr\gamma_{5}$ without any suppression.
In this sense, the relation $Tr\gamma_{5}=0$ is valid only when
we allow the presence of unphysical states in the Hilbert space;
this statement is also valid in the Pauli-Villars regularization.
If one  eliminates the contributions of the  unpysical states, 
the trace $Tr\gamma_{5}$ is replaced by $Tr\Gamma_{5}\equiv Tr \gamma_{5}(1 -\frac{1}{2}aD)$ which 
gives rise to the Pontryagin index, 
to be consistent with the continuum analysis.
\end{abstract}

\section{Introduction}
Recent developments in the treatment of  fermions in lattice gauge
theory led  to a better understanding of chiral symmetry[1]-[9].
 These developments are based on a  
 lattice Dirac  operator $D$ which satisfies  the so-called Ginsparg-Wilson relation[1]
\begin{equation}
\gamma_{5}D + D\gamma_{5} = aD\gamma_{5}D
\end{equation}
where  $\gamma_{5}$ is a hermitian chiral Dirac matrix.
 An explicit example of the operator satisfying (1.1) and free of species doubling has been given 
by Neuberger[2]. The operator has also been discussed as a fixed point  form of block 
transformations [3].
The relation (1.1) led to the  interesting analyses of the notion of  index in lattice gauge 
theory[4]-[9]. 

The index relation is often written as [4][5]
\begin{equation}
 - Tr\gamma_{5}\frac{1}{2}aD = n_{+} - n_{-}
\end{equation}
where $n_{\pm}$ stand for the number of  normalizable zero 
modes in 
\begin{equation}
D\varphi_{n}=0
\end{equation}
with simultaneous eigenvalues $\gamma_{5}\varphi_{n}= \pm \varphi_{n}$.  

 In the continuum  path integral treatment of chiral anomaly, 
the relation
\begin{equation}
Tr \gamma_{5}  = n_{+} - n_{-}
\end{equation}
in a suitably regularized sense (with help of a local version of
the index theorem[10]) plays a fundamental role[11]. On the other hand,
it is expected that the relation
\begin{equation}
Tr \gamma_{5} = 0
\end{equation}
holds on a finite lattice. As  Chiu correctly pointed out[12], 
this relation (1.5) leads to 
an interesting {\em constraint} 
\begin{equation}
Tr \gamma_{5} = n_{+}- n_{-} + N_{+} - N_{-} = 0  
\end{equation} 
where $N_{\pm}$ stand for the number of eigenstates $D\varphi_{n}=(2/a)\varphi_{n}$  with  
$\gamma_{5}\varphi_{n}= \pm \varphi_{n}$, respectively.

On the basis of the work [13], we here argue that 
$  Tr \gamma_{5} = 0$ implies the inevitable 
contribution from unphysical (would-be) species doublers in lattice theory or 
an unphysical bosonic spinor in Pauli-Villars regularization. In other
words, $Tr \gamma_{5} = 0$ cannot hold in the physical Hilbert space
consisting of physical states only, and the continuum limit of $  Tr
\gamma_{5} = 0$ is not defined consistently, as is seen in (1.6).  It is
also shown that the failure of the decoupling of heavy fermions in the anomaly calculation is 
crucial to
understand the consistency of  the customary lattice calculation of anomaly where $  Tr \gamma_{5} 
= 0$ is used.
( The continuum limit in this paper stands for the so-called ``naive''
continuum limit with $a\rightarrow 0$, and the lattice size is gradually extended to
infinity for any finite $a$ in the process of taking the limit $a\rightarrow 0$.)

\section{Finite dimensional representations of the Ginsparg - Wilson algebra}
We start with  the finite dimensional representations[13] of the basic algebraic relation (1.1). A 
construction of the operator $D$, which satisfies the Ginsparg-Wilson relation on a finite lattice, 
by using  a corresponding  operator $D$ on an 
infinite lattice has been discussed in Ref.[14]. We first  define an operator
\begin{equation}
\Gamma_{5}\equiv \gamma_{5}(1-\frac{1}{2}aD)
\end{equation}
which is hermitian and satisfies the basic relation derived from
(1.1)
\begin{equation}
\Gamma_{5}\gamma_{5}D +  \gamma_{5}D\Gamma_{5}=0.
\end{equation}
Namely, $\Gamma_{5}$ plays a role of $\gamma_{5}$ in continuum
theory. This relation suggests that if
\begin{equation}
\gamma_{5}D\phi_{n} = \lambda_{n}\phi_{n}, \ \ \ (\phi_{n},\phi_{n}) =1 
\end{equation}
for the {\em hermitian} operator $\gamma_{5}D$, then
\begin{equation}
\gamma_{5}D(\Gamma_{5}\phi_{n}) = -\lambda_{n}(\Gamma_{5}\phi_{n}).
\end{equation}
Namely, the eigenvalues $\lambda_{n}$ and $-\lambda_{n}$ are always paired 
if $\lambda_{n}\neq 0$ and $(\Gamma_{5}\phi_{n},\Gamma_{5}\phi_{n})\neq 0$. The inner product 
$\phi^{\dagger}_{n}\phi_{n} = (\phi_{n},\phi_{n})
\equiv \sum_{x} a^{4}\phi^{\star}_{n}(x)\phi_{n}(x)$ is defined by summing over all the lattice 
points, which are not explicitly written
in $\phi_{n}$.

We  evaluate the norm of $\Gamma_{5}\phi_{n}$
\begin{eqnarray}
(\Gamma_{5}\phi_{n},\Gamma_{5}\phi_{n})
&=& (\phi_{n},(\gamma_{5} - \frac{a}{2}\gamma_{5}D)(\gamma_{5} - 
\frac{a}{2}\gamma_{5}D)\phi_{n})\nonumber\\ 
&=&(\phi_{n},(1-  \frac{a}{2}\gamma_{5}(\gamma_{5}D+ D\gamma_{5}) + 
\frac{a^{2}}{4}(\gamma_{5}D)^{2})\phi_{n})\nonumber\\
&=&(\phi_{n},(1-  \frac{a^{2}}{4}(\gamma_{5}D)^{2})\phi_{n})\nonumber\\
&=&(1-  \frac{a}{2}\lambda_{n})(1+ \frac{a}{2}\lambda_{n}).
\end{eqnarray}
Namely $\phi_{n}$ is a ``highest'' state 
\begin{equation}
\Gamma_{5}\phi_{n}=(\gamma_{5} - \frac{a}{2}\gamma_{5}D)\phi_{n}=0
\end{equation}
if $(1-  \frac{a}{2}\lambda_{n})(1+ \frac{a}{2}\lambda_{n})=0 $
for the Euclidean $SO(4)$- invariant positive definite inner product 
$(\phi_{n}, \phi_{n})$.
We thus conclude that the states $\phi_{n}$ with $\lambda_{n}= \pm \frac{2}{a}$
 are {\em not} paired by the operation $\Gamma_{5}\phi_{n}$ and are the 
simultaneous eigenstates of $\gamma_{5}$, $\gamma_{5}\phi_{n}= \pm \phi_{n} $
respectively. These eigenvalues $\lambda_{n}$ are also the 
maximum or minimum of the possible eigenvalues of $\gamma_{5}D$. This is based
on the relation  
\begin{equation}
\phi^{\dagger}_{n}\gamma_{5}\phi_{n} = \frac{a}{2}\lambda_{n}\phi^{\dagger}_{n}\phi_{n} = 
\frac{a}{2}\lambda_{n}
\end{equation}
for $\lambda_{n}\neq 0$, which is derived by sandwiching the
relation(1.1) by $\phi^{\dagger}_{n}\gamma_{5}$ and $\phi_{n}$.Namely,  $ |\frac{a\lambda_{n}}{2}| 
= |\phi^{\dagger}_{n}\gamma_{5}\phi_{n}|\leq ||\phi_{n}|| ||\gamma_{5}\phi_{n}|| = 1 $.

For the vanishing eigenvalues $\gamma_{5}D\phi_{n} = 0$, one can
show that $\gamma_{5}D\gamma_{5}\phi_{n} = 0$ by using the 
relation (1.1). Thus one can choose the chiral eigenstates
$\gamma_{5}D[(1\pm \gamma_{5})/2]\phi_{n} = 0$. Namely
\begin{equation}
\gamma_{5}D\phi_{n} = 0, \ \ \ \gamma_{5}\phi_{n}=\pm\phi_{n}.
\end{equation}

All the remaining normalizable states with $0 < |\lambda_{n}| < 2/a$, which appear pairwise with 
$\lambda_{n}= \pm |\lambda_{n}|$ ( note that $\Gamma_{5}(\Gamma_{5}\phi_{n}) = (1 - 
(a\lambda_{n}/2)^{2})\phi_{n} \propto \phi_{n}$ for 
$|a\lambda_{n}/2|\neq 1$ )
, satisfy the relations
\begin{eqnarray}
\phi^{\dagger}_{n}\Gamma_{5}\phi_{n}&=&0,\nonumber\\
\phi^{\dagger}_{n}\gamma_{5}\phi_{n}&=& \frac{a\lambda_{n}}{2},\nonumber\\
\phi^{\dagger}_{m}\gamma_{5}\phi_{n}&=& 0 \ \ for \ \ \lambda_{m}\neq \lambda_{n}, \ \ 
\lambda_{m}\lambda_{n}>0.
\end{eqnarray}
These states $\phi_{n}$ cannot be the eigenstates of $\gamma_{5}$ as 
$|a\lambda_{n}/2| < 1$. 

The index (1.2) in the present context is most naturally defined as  
\begin{eqnarray}
Tr\Gamma_{5} &=& \sum_{n} \phi^{\dagger}_{n}\Gamma_{5}\phi_{n}\nonumber\\
&=& \sum_{ \lambda_{n}=0}\phi^{\dagger}_{n}\Gamma_{5}\phi_{n} +
\sum_{0<|\lambda_{n}|<2/a}\phi^{\dagger}_{n}\Gamma_{5}\phi_{n}
+\sum_{\lambda_{n}=\pm 2/a}\phi^{\dagger}_{n}\Gamma_{5}\phi_{n} 
\nonumber\\
&=& \sum_{ \lambda_{n}=0}\phi^{\dagger}_{n}\gamma_{5}\phi_{n}
= n_{+} - n_{-}
\end{eqnarray}
On the other hand, the index relation  commonly written in the form [4]-[6]
\begin{eqnarray}
 Tr( \frac{-1}{2}a\gamma_{5}D) &=& - \sum_{n}\phi_{n}^{\dagger}
\frac{a\gamma_{5}}{2}D\phi_{n}\nonumber\\
&=& -\frac{a}{2}\sum_{n}\lambda_{n}= -N_{+} + N_{-}
\end{eqnarray}
is saturated by the states $N_{\pm}$,where $N_{\pm}$ stand for the number of isolated (un-paired) 
states with $\lambda_{n}=
\pm 2/a$ and $\gamma_{5}\phi_{n} = \pm \phi_{n}$, respectively.

The relation $Tr \gamma_{5}=0$, which is expected to be valid on a finite lattice,  leads to ( by 
using (2.7))
\begin{eqnarray}
Tr\gamma_{5} &=& \sum_{n} \phi^{\dagger}_{n}\gamma_{5}\phi_{n}\nonumber\\
&=& \sum_{ \lambda_{n}=0}\phi^{\dagger}_{n}\gamma_{5}\phi_{n} +
\sum_{ \lambda_{n}\neq 0}\phi^{\dagger}_{n}\gamma_{5}\phi_{n}\nonumber\\
&=& \sum_{ \lambda_{n}=0}\phi^{\dagger}_{n}\gamma_{5}\phi_{n}
+ \sum_{ \lambda_{n}\neq 0}\frac{a}{2}\lambda_{n}\nonumber\\
&=& n_{+} - n_{-} +  \sum_{ \lambda_{n}\neq 0}\frac{a}{2}\lambda_{n}=0.
\end{eqnarray}
In the last line  of this relation, all the states except for the 
states with $\lambda_{n}= \pm 2/a$  cancel pairwise for $\lambda_{n}\neq
0$. We thus obtain a chirality  sum rule $ n_{+} - n_{-} + N_{+} -  N_{-} =0 
$ [12] or,  
\begin{equation}
n_{+}+ N_{+} =  n_{-} + N_{-}.  
\end{equation} 
 These relations show that the chirality asymmetry at vanishing eigenvalues is balanced by the 
chirality asymmetry at the largest  eigenvalues with $|\lambda_{n}|= 2/a$. 

Those properties we analyzed so far in this Section hold both for non-Abelian and Abelian gauge 
theories. We did not specify precise boundary conditions, since our analysis is valid once 
non-trivial zero modes appear for a given boundary condition. For an Abelian theory, one needs to 
introduce the gauge field configuration  with suitable boundary conditions, which carries a  
non-vanishing magnetic flux , to generate a non-trivial index $n_{+} - n_{-}$ [14]. Our analysis of 
the index in this Appendix  is  formal, since it is well known that 
the Ginsparg-Wilson relation (1.1) by itself does not uniquely specify
the index or the coefficient of chiral anomaly for a given gauge field
configuration [15].A suitable choice of parameters in the 
overlap operator [2], for example, uniquely specifies the index.  

To summarize the analyses of the present Section, all the normalizable
eigenstates $\phi_{n}$ of the hermitian  $\gamma_{5}D$ on a finite lattice are categorized into the 
following 3 classes:\\
(i)\ $n_{\pm}$ states,\\
\begin{equation}
\gamma_{5}D\phi_{n}=0, \ \ \gamma_{5}\phi_{n} = \pm \phi_{n},
\end{equation}
(ii)\ $N_{\pm}$ states($\Gamma_{5}\phi_{n}=0$), \\
\begin{equation}
\gamma_{5}D\phi_{n}= \pm \frac{2}{a}\phi_{n}, \ \ \gamma_{5}\phi_{n} = \pm \phi_{n},\ \ 
respectively,
\end{equation}
(iii) Remaining states with $0 < |\lambda_{n}| < 2/a$,
\begin{equation}
\gamma_{5}D\phi_{n}= \lambda_{n}\phi_{n}, \ \ \ \gamma_{5}D(\Gamma_{5}\phi_{n})
= - \lambda_{n}(\Gamma_{5}\phi_{n}), 
\end{equation}
and the sum rule $n_{+}+ N_{+} =  n_{-} + N_{-}$ holds. 

All the $n_{\pm}$ and 
$N_{\pm}$ states are the eigenstates of $D$,  $D\phi_{n}=0$ and $D\phi_{n}= (2/
a) \phi_{n}$, respectively. If one denotes the number of states  in (iii) 
by $2N_{0}$, the total number of states (dimension of the 
representation) $N$ is given by $N = 2(n_{+} + N_{+} +
N_{0})$, which is expected to be a constant independent of background gauge 
field configurations.

\section{The nature of the $N_{\pm}$ states with $\Gamma_{5}\phi_{n} = 0$ }
In the previous section we have seen  that the consistency of the
relation $Tr \gamma_{5} = 0$ requires the presence of the $N_{\pm}$
states for an operator $\gamma_{5}D$ satisfying (1.1) on a finite
lattice. We want to show that the $N_{\pm}$ states are 
topological excitations associated with species doublers. For this purpose, we start with the 
conventional Wilson operator $D_{W}$
\begin{eqnarray}
D_{W}(n,m)&\equiv&i\gamma^{\mu}C_{\mu}(n,m) + B(n,m) -\frac{1}{a}m_{0}\delta_{n,m},\nonumber\\
C_{\mu}(n,m)&=&\frac{1}{2a}[\delta_{m+\mu,n}U_{\mu}(m) - 
\delta_{m,n+\mu}U^{\dagger}_{\mu}(n)],\nonumber\\
B(n,m)&=&\frac{r}{2a}\sum_{\mu}[2\delta_{n,m}-\delta_{m+\mu,n}U_{\mu}(m)
-\delta_{m,n+\mu}U^{\dagger}_{\mu}(n)],\nonumber\\
U_{\mu}(m)&=& \exp [iagA_{\mu}(m)],
\end{eqnarray}
where we added a constant mass term to $D_{W}$. Our 
matrix convention is that $\gamma^{\mu}$ are anti-hermitian, $(\gamma^{\mu})^{\dagger} = - 
\gamma^{\mu}$, and thus $\not \!\! C\equiv \gamma^{\mu}C_{\mu}(n,m)$ is hermitian
\begin{equation}
\not \!\! C ^{\dagger} = \not \!\! C.
\end{equation}

In the case of  $\Cslash$, a very explicit construction of species doublers 
is known.  For a square lattice one can explicitly show that the simplest lattice fermion action
\begin{equation}
S = \bar{\psi}i\Cslash\psi
\end{equation}
is invariant under the transformation[16]
\begin{equation}
\psi^{\prime}= {\cal T}\psi,\ \bar{\psi}^{\prime}= \bar{\psi}{\cal T}^{-1}
\end{equation}
where ${\cal  T}$ stands for any one of the following 16 operators
\begin{equation}
1,\ T_{1}T_{2},\ T_{1}T_{3},\ T_{1}T_{4},\ T_{2}T_{3},\ T_{2}T_{4},\ T_{3}T_{4},\ 
T_{1}T_{2}T_{3}T_{4},
\end{equation}
and 
\begin{equation}
T_{1},\ T_{2},\ T_{3},\ T_{4},\ T_{1}T_{2}T_{3},\ T_{2}T_{3}T_{4},\ T_{3}T_{4}T_{1},\ 
T_{4}T_{1}T_{2}.
\end{equation}
The operators  $T_{\mu}$  are  defined by 
\begin{equation}
T_{\mu}\equiv \gamma_{\mu}\gamma_{5}\exp {(i\pi x^{\mu}/a)}  
\end{equation}
and  satisfy the relation
\begin{equation}
T_{\mu}T_{\nu} + T_{\nu}T_{\mu}=2\delta_{\mu\nu}
\end{equation}
with  $T_{\mu}^{\dagger} = T_{\mu} = T^{-1}_{\mu}$ for anti-hermitian $\gamma_{\mu}$. 
We denote the 16 operators by ${\cal T}_{n}, \ \ n=0\sim 15$, in the following 
with ${\cal T}_{0}=1$.
By recalling that the operator $T_{\mu}$ adds the  momentum $\pi/a$ to the 
fermion momentum $k_{\mu}$, we cover the entire Brillouin zone  
\begin{equation}
- \frac{\pi}{2a} \leq k_{\mu} <  \frac{3\pi}{2a}
\end{equation}
by the operation (3.4) starting with the free fermion defined in
\begin{equation}
- \frac{\pi}{2a} \leq k_{\mu} <  \frac{\pi}{2a}.
\end{equation}
The operators in (3.5) commute with $\gamma_{5}$, whereas those in (3.6) anti-commute with 
$\gamma_{5}$ and thus change the sign of  chiral charge,  reproducing the 15 species doublers with 
correct chiral charge assignment; $\sum_{n=0}^{15}(-1)^{n}\gamma_{5} =0$. 

In a smooth continuum limit, the operaor $\Cslash$ produces  
$\Dslash$ for each species doubler with alternating chiral charge. The relation $Tr \gamma_{5} = 0$ 
for the operator $\Cslash$ is consistent  for any background gauge field because of the presence of 
these species doublers, which are degenerate with the physical species in the present case.

The consistency of $Tr\gamma_{5} = 0$ is generally analyzed by means of topological properties  and 
thus it is best described in the nearly continuum limit. 
To be more precise, one may define  the near continuum configurations by
the momentum $k_{\mu}$ carried by the fermion
\begin{equation}
- \frac{\pi}{2a}\epsilon \leq k_{\mu} \leq \frac{\pi}{2a}\epsilon
\end{equation}
for sufficiently small $a$ and $\epsilon$ combined with the operation ${\cal T}_{n}$ in (3.5) and 
(3.6). 
To identify each species doubler clearly in the near continuum configurations, we also keep $r/a$ 
and $m_{0}/a$ finite for $a\rightarrow$ small [16], and the gauge fields are assumed to be 
sufficiently smooth. For these configurations, we can approximate the operator $D_{W}$ by
\begin{equation}
D_{W}= i\Dslash + M_{n} + O(\epsilon^{2}) + O(agA_{\mu})
\end{equation}
for each species doubler, where the mass parameters $M_{n}$  stand for $M_{0}= - \frac{m_{0}}{a}$ 
and one of 
\begin{eqnarray}
&&\frac{2r}{a}-\frac{m_{0}}{a},\ \ (4,-1);\ \ \ 
\frac{4r}{a}-\frac{m_{0}}{a},\ \ (6,1)\nonumber\\
&&\frac{6r}{a}-\frac{m_{0}}{a},\ \ (4,-1);\ \ \ 
\frac{8r}{a}-\frac{m_{0}}{a},\ \ (1,1)
\end{eqnarray}
for $n=1\sim 15$[2]. Here we denoted ( multiplicity, chiral charge ) in the bracket for species 
doublers. In (3.12) we used the relation valid for the configurations (3.11), for example, 
\begin{eqnarray}
D_{W}(k) &=& \sum_{\mu}\gamma^{\mu}\frac{\sin ak_{\mu}}{a} + \frac{r}{a}\sum_{\mu}(1 - \cos 
ak_{\mu}) - \frac{m_{0}}{a}\nonumber\\
&=& \gamma^{\mu}k_{\mu}( 1 + O(\epsilon^{2})) + \frac{r}{a} O(\epsilon^{2}) -
\frac{m_{0}}{a}
\end{eqnarray}
in the momentum representation with vanishing gauge field. 

In these near continuum configurations, the topological properties are specified by the operator  
$\Dslash$  in $D_{W}$.
The overlap operator $D$ introduced by Neuberger[2] , which satisfies the relation (1.1), has an 
explicit expression
\begin{equation}
aD= 1 - \gamma_{5}\frac{H}{\sqrt{H^{2}}} =1 + D_{W}\frac{1}{\sqrt{D_{W}^{\dagger}D_{W}}}
\end{equation}
where $D_{W}=-\gamma_{5} H$ is the Wilson operator.
For the near continuum configurations specified above in (3.11), one can approximate 
\begin{eqnarray}
D&=& \sum_{n=0}^{15}(1/a){[}1 + (i\Dslash + M_{n})\frac{1}{\sqrt{\Dslash^{2} + 
M_{n}^{2}}}]|n\rangle\langle n|,\nonumber\\
\gamma_{5}D&=& \sum_{n=0}^{15}(-1)^{n}\gamma_{5}(1/a){[}1 + (i\Dslash + 
M_{n})\frac{1}{\sqrt{\Dslash^{2} + M_{n}^{2}}}]|n\rangle\langle n|,\nonumber\\ 
\gamma_{5}&=& \sum_{n=0}^{15}(-1)^{n}\gamma_{5}|n\rangle\langle n|.
\end{eqnarray}
Here we explicitly write the projection  $|n\rangle\langle n|$  for each species doubler. The 
operators in (3.16) preserve the Ginsparg-Wilson relation (1.1).

The above expression of $D$ in (3.16) shows that 
\begin{eqnarray}
D\phi_{l}& =& 0, \nonumber\\
D\phi_{l}&=& \frac{2}{a}\phi_{l}
\end{eqnarray}
for the physical species and the unphysical species doublers,
respectively, if one uses the {\em zero-modes} in
\begin{equation}
\Dslash\phi_{l}=\lambda_{n}\phi_{n}.
\end{equation}  
Note that $M_{0}<0$ and 
the rest of $M_{n}>0$ in (3.13) and (3.16) [2]. We also note that $\phi_{l}$
can be a simultaneous eigenstate of $\gamma_{5}$ only for $\Dslash\phi_{l}=0$.  Namely, the 
$N_{\pm}$ states
with eigenvalue $2/a$ of $D$ in fact correspond to {\em topological 
excitations} associated with  species doublers. This means that the multiplicities of these 
$N_{\pm}$ are quite high  due to the 15 species 
doublers, although they  satisfy the sum rule  $n_{+} + N_{+} = n_{-} + N_{-}$:
This sum rule itself is a direct consequence of (3.17), (3.18)
and $\sum_{n}(-1)^{n}\gamma_{5}=0$ if one recalls
$\phi_{l}^{\dagger}\gamma_{5}\phi_{l}= 0$ for $\lambda_{l}\neq 0$. The
relation  (3.17) with zero-modes in (3.18) shows that the index 
relation (2.10) stands for a lattice version of the Atiyah-Singer
index theorem. 

 We here briefly sketch an evaluation of the anomaly, since it  
justifies our analysis based 
on the effective expressions in (3.16). For an operator $O(x,y)$ defined on the lattice,
one may define 
\begin{equation}
O_{mn}\equiv \sum_{x,y}\phi_{m}^{\ast}(x)O(x,y)\phi_{n}(y),
\end{equation}
and the trace
\begin{eqnarray}
Tr O &=& \sum_{n}O_{nn}\nonumber\\
&=&\sum_{n}\sum_{x,y}\phi_{n}^{\ast}(x)O(x,y)\phi_{n}(y)\nonumber\\
&=&\sum_{x}(\sum_{n,y}\phi_{n}^{\ast}(x)O(x,y)\phi_{n}(y)).
\end{eqnarray}
The local version of the trace (or anomaly) is then defined by $tr O(x,x) \equiv\\ 
\sum_{n,y}\phi_{n}^{\ast}(x)O(x,y)\phi_{n}(y)$. For the operator of our interest, we have 
\begin{equation}
tr\gamma_{5} (1-\frac{a}{2}D)(x) =
-\frac{1}{2}\sum_{n=0}^{15} tr 
\int^{\frac{\pi}{2a}}_{-\frac{\pi}{2a}}\frac{d^{4}k}{(2\pi)^{4}}e^{-ikx}{\cal T}^{-1}_{n}\gamma_{5}
\frac{D_{W}}{\sqrt{D_{W}^{\dagger}D_{W}}}{\cal T}_{n}e^{ikx}
\end{equation}
where we used the plane wave basis defined in (3.10) combined with the operation
${\cal T}_{n}$. We also used the chiral charge assignment (3.16)
and a short hand notation $Oe^{ikx}= \sum_{y}O(x,y)e^{iky}$.

We first take the  $a\rightarrow 0$ limit of (3.21) with all $M_{n}, 
n=0\sim 15$, kept fixed and then take the limit $|M_{n}| \rightarrow \infty$
later. For fixed $M_{n}$ ( to be precise, for fixed $m_{0}/a$ and $r/a$ ),
one can confirm that the above integral (3.21) for the domain $\frac{\pi}{2a}\epsilon \leq |k_{\mu}
| \leq \frac{\pi}{2a}$ vanishes ( at least ) linearly in $a$ for $a \rightarrow 0$, if one takes 
into account the trace with $\gamma_{5}$. See also Refs.[7][9]. In the remaining integral 
\begin{equation}
-\frac{1}{2}\sum_{n=0}^{15} (-1)^{n}tr 
\int^{\frac{\pi}{2a}\epsilon}_{-\frac{\pi}{2a}\epsilon}\frac{d^{4}k}{(2\pi)^{4}}e^{-ikx}\gamma_{5}{
\cal T}^{-1}_{n}D_{W}\frac{1}{\sqrt{D_{W}^{\dagger}D_{W}}}{\cal T}_{n}e^{ikx}
\end{equation}
one may take the limit $a\rightarrow 0$ ( and $\frac{\pi}{2a}\epsilon \rightarrow \infty$ ) with 
letting $\epsilon$ arbitrarily small. By taking (3.12) into 
account, one thus recovers the local anomaly [8]
\begin{eqnarray}
&& - \frac{1}{2} \sum_{n=0}^{15}(-1)^{n}tr \int^{\infty}_{-\infty} \frac{d^{4}k}{(2\pi)^{4}} 
e^{-ikx}\gamma_{5}(i\Dslash + M_{n})\frac{1}{\sqrt{\Dslash^{2} + M_{n}^{2}}}e^{ikx}\nonumber\\
&=&\frac{1}{2}tr \int^{\infty}_{-\infty} \frac{d^{4}k}{(2\pi)^{4}} 
e^{-ikx}\gamma_{5}[\frac{1}{\sqrt{\Dslash^{2}/M_{0}^{2}+1}}- \sum_{n=1}^{15}(-1)^{n} 
\frac{1}{\sqrt{\Dslash^{2}/M_{n}^{2}+1}}]e^{ikx}\nonumber\\
&=& \frac{g^{2}}{32\pi^{2}}tr \epsilon^{\mu\nu\alpha\beta}F_{\mu\nu}F_{\alpha\beta} 
\end{eqnarray}
for $|M_{n}|\rightarrow \infty$. The domain in (3.11) with arbitrarily small but finite $\epsilon$ 
thus correctly describes the topological aspects of the continuum limit in the present 
prescription. 

Here we went through some  details of the anomaly calculation to show that the interpretation of 
the 
$N_{\pm}$ states in (2.15) as topological excitations associated 
with 
species doublers, as is shown in (3.17), is also consistent with the  local 
anomaly calculation. As for a general analysis of chiral anomaly
in the overlap operator, see Ref.[17].
 
\subsection{General lattice Dirac operator and $Tr \gamma_{5}=0$}
We expect that our analysis of $Tr \gamma_{5}=0$, namely its consistency
is ensured only by the presence of the would-be species doublers in the
Hilbert space,  works for a general lattice Dirac operator, since any
lattice operator contains $\Cslash$ as an essential part. For the smooth
near continuum configurations, the lowest dimensional operator $\Cslash$
is expected to specify the topological properties.

We also note that the Pauli-Villars regularization in continuum theory can be 
analyzed in a similar way. The Pauli-Villars regulator is defined in the path 
integral by introducing a bosonic spinor $\phi$ into the action
\begin{equation}
S = \int d^{4}x [\bar{\psi}(i\Dslash - m )\psi + \bar{\phi}(i\Dslash - M )
\phi ].
\end{equation}
The Jacobian for the global chiral transformation then gives rise to the graded
trace[11]
\begin{equation}
Tr \gamma_{5} = Tr_{\psi}\gamma_{5} - Tr_{\phi}\gamma_{5} = 0.
\end{equation}
The relation $Tr \gamma_{5} = 0$ is thus consistent with any topologically non-trivial background  
gauge field because of the presence of the unphysical regulator $\phi$. This $\phi$ is analogous to 
the species doublers in lattice regularization. 

\section{Implications of the present analysis}
We have shown that the consistency of $Tr \gamma_{5} = 0$ for topologically non-trivial background 
gauge fields requires the presence
 of some unphysical states in the Hilbert space. Coming back to the original lattice theory defined 
by 
\begin{equation}
S = \bar{\psi}D\psi
\end{equation}
with $D$ satisfying the relation (1.1), one obtains twice of 
 (2.10) as a Jacobian factor for the global chiral transformation [5] $\delta\psi = 
i\epsilon\gamma_{5}(1-\frac{a}{2}D)\psi$ and $\delta\bar{\psi} = \bar{\psi}i\epsilon (1-\frac{a}{2}
D)\gamma_{5}$, which leaves the action (4.1) invariant. One can rewrite (2.10) as 
\begin{equation}
Tr \gamma_{5}( 1 - \frac{a}{2}D) = \tilde{T}r \gamma_{5}( 1 - \frac{a}{2}D) =
\tilde{T}r \gamma_{5} = n_{+} - n_{-}
\end{equation}
where the modified trace $\tilde{T}r$ is defined  by truncating the unphysical 
$N_{\pm}$ states with $\lambda_{n} =\pm  2/a$ . Without the $N_{\pm}$ states, $\tilde{T}r 
\gamma_{5}\frac{a}{2}D = 0$ since the eigenvalues $\lambda_{n}$ of $\gamma_{5}D$ with $\lambda_{n} 
\neq 0, \pm 2/a$ appear always pairwise at $\pm |\lambda_{n}|$. 

If one takes a smooth continuum limit of $\tilde{T}r \gamma_{5}= n_{+} - n_{-}$ in (4.2) , one 
recovers the result of the continuum path integral (1.4). If one considers that $\tilde{T}r 
\gamma_{5}$ is too abstract, one may  define it more concretely by
\begin{eqnarray}
Tr \gamma_{5}( 1 - \frac{a}{2}D) f(\frac{(\gamma_{5}D)^{2}}{M^{2}})&=& \tilde{T}r \gamma_{5}( 1 - 
\frac{a}{2}D)f(\frac{(\gamma_{5}D)^{2}}{M^{2}})\nonumber\\
&=& \tilde{T}r \gamma_{5}f(\frac{(\gamma_{5}D)^{2}}{M^{2}}) = n_{+} - n_{-}
\end{eqnarray}
for {\em any } $f(x)$ which rapidly goes to $0$ for 
$x \rightarrow\infty$ with $f(0)=1$.  This relation 
suggests that we can extract the local index [10]
( or anomaly) by  
\begin{equation}
tr \gamma_{5}( 1 - \frac{a}{2}D) f(\frac{(\gamma_{5}D)^{2}}{M^{2}})(x)  
\end{equation}
which  is shown to be independent of the choice of $f(x)$ in the limit
$a \rightarrow 0$ and leads to (3.23) by using only the general
properties of $D$ [8].  We thus naturally recover the result of the
continuum path integral[11]. As for an interesting algebraic 
analysis of the local anomaly $tr \gamma_{5}( 1 - \frac{a}{2}D)$,
see [18].

As for a more practical implication of $Tr \gamma_{5} = 0$ in  lattice theory, one may say that any  
result which depends  explicitly on the presence of the states $N_{\pm}$  is  {\em unphysical}. It 
is thus 
necessary to define the scalar density ( or mass term ) and pseudo-scalar
density in the theory (4.1) by [19][6]
\begin{eqnarray}
S(x) &=& \bar{\psi}_{L}\psi_{R} + \bar{\psi}_{R}\psi_{L}= \bar{\psi}( 1 - 
\frac{a}{2}D)\psi,\nonumber\\
P(x) &=& \bar{\psi}_{L}\psi_{R} - \bar{\psi}_{R}\psi_{L}= \bar{\psi}\gamma_{5}( 1 - 
\frac{a}{2}D)\psi.
\end{eqnarray}
Here we defined two independent projection operators 
\begin{eqnarray}
P_{\pm} &=& \frac{1}{2}( 1 \pm \gamma_{5})\nonumber\\
\hat{P}_{\pm} &=& \frac{1}{2}( 1 \pm \hat{\gamma}_{5})   
\end{eqnarray}
with $\hat{\gamma}_{5} = \gamma_{5}(1 - aD)$ which satisfies $\hat{\gamma}_{5}^{2} = 1$ [6]. The 
left- and right- components are then defined by 
\begin{equation}
\bar{\psi}_{L,R}= \bar{\psi}P_{\pm}, \ \ \psi_{R,L}= \hat{P}_{\pm}\psi
\end{equation}
which is based on the decomposition
\begin{equation}
D = P_{+}D\hat{P}_{-} +  P_{-} D\hat{P}_{+}.    
\end{equation}
The physical operators $S(x)$ and $P(x)$ in (4.5) do not contain the contribution from the 
unphysical
states $N_{\pm}$ in (2.15). In the spirit of  this construction, the definition of the index  by 
(4.3) which is independent of unphysical states $N_{\pm}$ is natural.
 In particular, all the unphysical species doublers ( not only the topological
ones at $2/a$ ) decouple from the anomaly defined by (4.4) in the limit 
$a \rightarrow 0$ with fixed $M$.  

The customary calculation of the index ( and also anomaly ) by 
using the relation[4]-[7][9]
\begin{equation}
Tr \gamma_{5}( 1 - \frac{a}{2}D) = Tr( - \frac{a}{2}\gamma_{5}D) = n_{+} - n_{-} 
\end{equation}
by itself is of course consistent, since one simply includes the unphysical states $N_{\pm}$ in 
evaluating  $Tr \gamma_{5} = 0$, and consequently one obtains the index 
$Tr ( - \frac{a}{2}\gamma_{5}D)$ from the unphysical states $N_{\pm}$ only.
We after all know that the left-hand side of (4.9) is independent of $N_{\pm}$. 

Rather, the major message of our analysis is that the continuum limit of 
$Tr \gamma_{5} = 0$ in (1.6) ( unlike the relation $\tilde{T}r
\gamma_{5} = n_{+} - n_{-}$ ) {\em cannot } be defined in a consistent
way when the (would-be) species doublers disappear from the Hilbert
space. It is clear from the expression of $Tr \gamma_{5} = 0$ in (1.6)
that the $a\rightarrow 0$ limit of $Tr \gamma_{5} = 0$ is not defined
consistently. One may then ask how the calculation of local anomaly on
the basis of (4.9) could be consistent in the limit $a\rightarrow 0$ if
$Tr \gamma_{5} = 0$ is inconsistent?  A key to resolve this apparent
paradox is the failure of the decoupling of heavy fermions in the
evaluation of anomaly. The massive unphysical species doublers do not
decouple from the anomaly , as is seen in (3.23), for example. If one
insists on $Tr \gamma_{5} = 0$ in the continuum limit, one is also
insisting on the failure of the decoupling of these infinitely massive
particles from $Tr \gamma_{5} = 0$. The contributions of these heavy
fermions to the anomaly and to $Tr \gamma_{5} = 0$ precisely cancel,
just as in the case of the evaluation of global index in (4.9). Namely,
the local anomaly itself is {\em independent} of these massive species
doublers in the  continuum limit, as is clear in (4.4). It is an
advantage of the finite lattice formulation that we can now clearly
illustrate this subtle cancellation of the contributions of those
ultra-heavy $N_{\pm}$-states  to  $Tr \gamma_{5} = 0$ and 
anomaly on the basis of (1.6).

The unphysical nature of the $N_{\pm}$-states has been recently
clarified from a different view point in Re.[20]

\end{document}